\documentclass[prl,twocolumn,showpacs,preprintnumbers,amsmath,amssymb,superscriptaddress,floatfix]{revtex4-2}

\usepackage{amsmath}    % need for subequations
\usepackage{amssymb,environ}
\usepackage{graphicx}   % need for figures
\usepackage{verbatim}   % useful for program listings
\usepackage{color}      % use if color is used in text
\usepackage{hyperref}   % use for hypertext links, including those to external documents and URLs
\usepackage{breakurl}
\usepackage[normalem]{ulem}
\usepackage{natbib}
\usepackage{enumitem}
\usepackage{multirow}

\usepackage{epsfig}
\usepackage{textcomp}
\usepackage{wasysym}
\usepackage{array}
\usepackage{color}

\begin{document}

\newcommand{\ev}[0]{\mathbf{e}}
\newcommand{\cv}[0]{\mathbf{c}}
\newcommand{\fv}[0]{\mathbf{f}}
\newcommand{\Rv}[0]{\mathbf{R}}
\newcommand{\Tr}[0]{\mathrm{Tr}}
\newcommand{\ud}[0]{\uparrow\downarrow}
\newcommand{\Uv}[0]{\mathbf{U}}
\newcommand{\Iv}[0]{\mathbf{I}}
\newcommand{\Hv}[0]{\mathbf{H}}

\setlength{\jot}{2mm}

\newcommand{\zg}[1]{{\color{red}#1}}
\newcommand{\red}[1]{{\color{red}#1}}
\newcommand{\ds}[1]{{\color{blue}#1}}

\newcommand{\dbra}[1]{\langle\!\langle #1}
\newcommand{\dket}[1]{|#1\rangle\!\rangle}

%\title{Anomalous self-energy in the one dimensional Hubbard model at infinite temperature}

\title{Anomalous Quantum Relaxation in the Infinite Temperature Hubbard Chain}

\author{C\u at\u alin Pa\c scu Moca}
\affiliation{Department of Theoretical Physics, Institute of Physics, Budapest University of Technology and Economics, M\H uegyetem rkp. 3., H-1111 Budapest, Hungary}
\affiliation{Department of Physics, University of Oradea,  410087, Oradea, Romania}

%\author{Gergely Zar\'and}
%\affiliation{Department of Theoretical Physics, Institute of Physics, Budapest University of Technology and Economics, M\H uegyetem rkp. 3., H-1111 Budapest, Hungary}

\author{Bal\'azs D\'ora}
\email{dora.balazs@ttk.bme.hu}
\affiliation{Department of Theoretical Physics, Institute of Physics, Budapest University of Technology and Economics, M\H uegyetem rkp. 3., H-1111 Budapest, Hungary}
\affiliation{MTA-BME Lend\"ulet "Momentum" Open Quantum Systems Research Group, Institute of Physics, Budapest University of Technology and Economics, 
M\H uegyetem rkp. 3., H-1111, Budapest, Hungary}

\begin{abstract}

The self-energy encodes the fundamental lifetime of quasiparticle excitations. In one dimension, it is 
known to display anomalous behavior at zero temperature for interacting fermions, reflecting the breakdown of Fermi-liquid 
theory. Here we show that the self-energy is also anomalous  in the infinite temperature Hubbard chain, where 
thermal fluctuations are maximal. Focusing on the second-order ring diagram, we find that the imaginary 
part of the self-energy diverges non-perturbatively: as a power law with exponent $-1/3$ near half 
filling, and logarithmically away from it. These divergences imply anomalous temporal relaxation of 
Green’s functions, confirmed by infinite temperature tensor-network simulations. Our results 
demonstrate that anomalous relaxation  and the breakdown of perturbation theory survive even at maximal 
entropy, which  can be observed in cold-atom experiments probing the  Hubbard chain at high temperatures.

% The self energy of interacting fermions in the one dimension
% has been long known to display anomalous behaviour at zero temperature and results in the breakdown of Fermi-liquid picture.
% Here we show that the very same quantity is also anomalous in the infinite temperature Hubbard model  by focusing on the second order ring diagram contribution to the self-energy.
% For the symmetric Hubbard model, 
% we find that the imaginary part of the self-energy diverges  on the mass shell in a power law fashion with exponent $-1/3$ for both energy and momentum close to half filling condition, 
% which changes into logarithmic behaviour 
% upon moving away from half filling.
% In the Falicov-Kimball limit, the self-energy is momentum independent but log divergent close to zero frequency for the immobile electrons  and displays an inverse 
% square root divergence for the mobile electrons at the band edge.
% Our analytical results are supported by infinite temperature numerical simulation of the Hubbard model.
\end{abstract}

\maketitle

\paragraph{Introduction.}

Infinite temperature quantum dynamics often\cite{ljubotina,PRX2025,Gopalakrishnan2023,rosenberg} reveals the surprising resilience of quantum phenomena even in the most extreme thermal conditions.
There,  many-body
systems reach a state of maximal entropy where all
microscopic configurations are equally populated.
One
might expect such conditions to suppress coherent quantum phenomena entirely, leaving only trivial classical
dynamics. However, recent developments have shown
that nontrivial correlations and anomalous dynamical
features can persist even at maximal thermal disorder. %, challenging the expectation that quantum effects vanish at high temperatures\cite{niemijer,sur,katsura}.
These non-classical features\cite{niemijer,sur,katsura} and eventually quantum coherence are uniquely probed by  the quasiparticle lifetime, which, at infinite temperature, is a diagnostic tool
of the underlying structure and dynamics in many-body systems, revealing
deep information about interactions, integrability, thermalization, and transport.

A natural setting to explore anomalous dynamics at infinite temperature  is the one-dimensional Hubbard chain, one of the most 
fundamental models of correlated electrons~\cite{arovas,essler2005,PhysRevX.5.041041}. Despite its 
deceiving simplicity, the model hosts rich physics, including spin-charge separation, non-Fermi liquid 
behavior, pseudogap, Mott insulating phases, and unconventional transport~\cite{giamarchi,mahan,Giuliani2005,Vanhala.2016}. While these features are well understood at low temperatures, far less is known about 
their fate at infinite temperature. Remarkably, recent work has shown that even in this extreme limit, 
spin and charge transport exhibits Kardar–Parisi–Zhang  scaling~\cite{kpz,moca2023}, 
suggesting that strong correlations and dynamical universality can survive heating to maximal entropy.

In this work, we investigate the single-particle Green’s function of the infinite temperature Hubbard 
chain, focusing on the role of interactions in shaping relaxation dynamics. By analytically evaluating 
the lowest-order ring diagram contribution to the self-energy, we demonstrate that it diverges on the 
mass shell, indicating a breakdown of perturbation theory. We identify a power-law divergence with 
exponent $-1/3$ near $k=\pm \pi/2$ and $\omega=0$, which crosses over to logarithmic behavior away from 
this point. These predictions are corroborated by infinite temperature tensor-network simulations of 
the Green’s function and self-energy. 
Our results establish that anomalous quantum relaxation is not a low-temperature 
peculiarity but a robust feature of the Hubbard chain even at infinite temperature, opening the way for 
experimental tests in cold-atom platforms.

\paragraph{The asymmetric Hubbard model.}

The asymmetric (or mass-imbalanced) Hubbard model~\cite{Farkasovsky,maska,wang2007,heitmann,jin2015} is defined as
\begin{gather}
H=-\sum_{j,\sigma} \frac{J_\sigma}{2} \left(c_{j\sigma}^\dagger c_{j+1\sigma}+ \text{h.c.}\right)+U\sum_j n_{j\uparrow}n_{j\downarrow},
\label{ahubbard}
\end{gather}
where $N$ sites are subject to periodic boundary conditions in one dimension, the hopping amplitudes are spin dependent, and $n_{j\sigma}=c_{j\sigma}^\dagger c_{j\sigma}$. 
The noninteracting dispersion is $\varepsilon_\sigma(k)=-J_\sigma \cos(k)$ with the lattice constant set to unity. This model interpolates continuously 
between the conventional Hubbard model ($J_\uparrow=J_\downarrow$) and the Falicov-Kimball limit\cite{falicov} ($J_\uparrow=0$, $J_\downarrow\neq 0$), exhibiting a rich phase diagram.

\begin{figure}[ht]
\centering
\includegraphics[width=4cm]{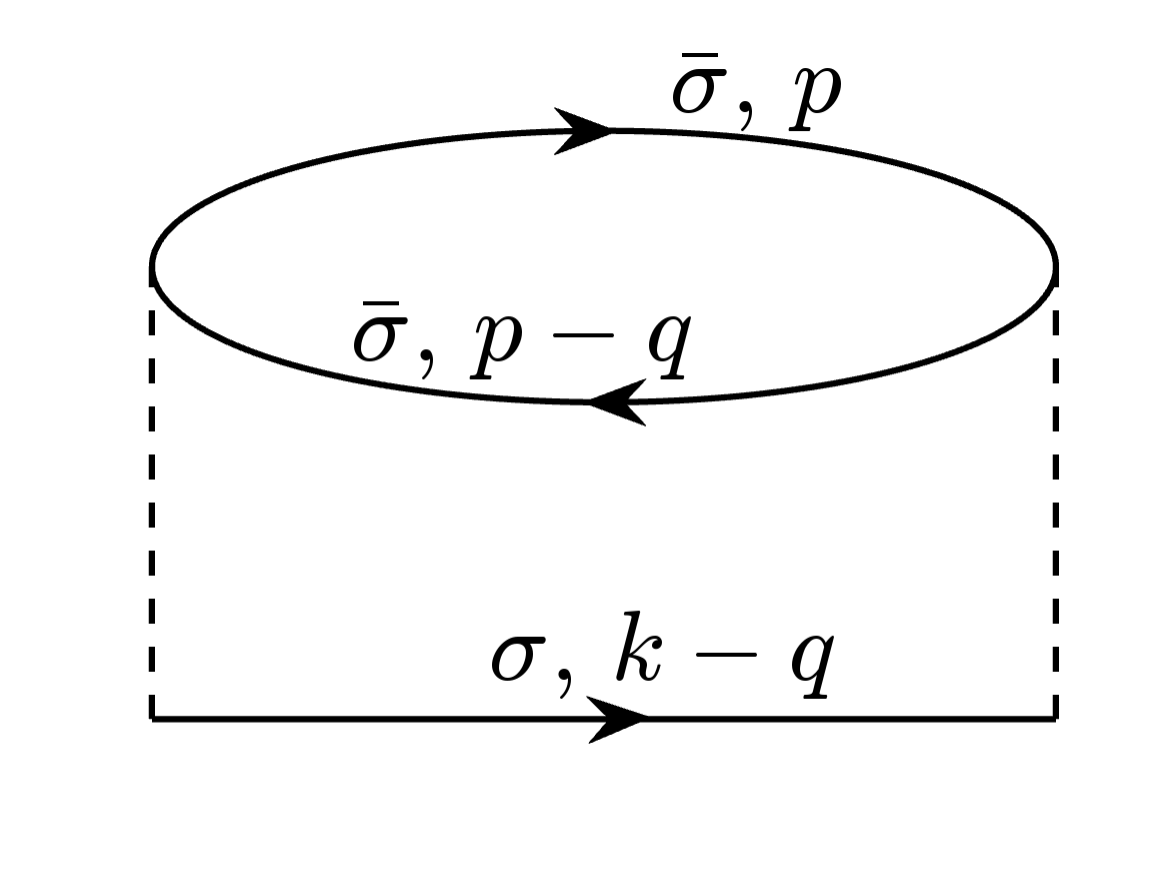}
\caption{The second order ring diagram to the self-energy is shown for electrons with spin $\sigma$. 
The solid/dashed lines correspond to electron Green's function/interaction, respectively, the spin and momentum are  explicitly
indicated.}
\label{ringdiag}
\end{figure}

Interactions enter the Green’s function through the self-energy~\cite{mahan,bruus,Giuliani2005},
$G_\sigma(k,\omega_n)=(i\omega_n-\varepsilon_\sigma(k)-\Sigma_\sigma(k,\omega_n))^{-1}$. Among the diagrammatic contributions, the second-order ring 
diagrams are known to dominate~\cite{Fetter,rhodes,zlatic91,kang1994,galan}, and we focus on their effect at infinite temperature (Fig.~\ref{ringdiag}). 
Performing the internal Matsubara sums in the high-temperature limit, we obtain
\begin{gather}
\Sigma_\sigma(k,\omega_n)= \sum_{p,q} \frac{U^2 \nu(1-\nu)/N^2}{i\omega_n-\varepsilon_{\bar\sigma}(p)+\varepsilon_{\bar\sigma}(p-q)-\varepsilon_\sigma(k-q)},
\label{self1}
\end{gather}
where $\bar\sigma=-\sigma$, $\nu$ is the spin-resolved filling, and $\omega_n=(2n+1)\pi T$ is a fermionic Matsubara frequency with $T$ the temperature. 
Analytic continuation $i\omega_n\to \omega+i0^+$ removes the temperature dependence, and the imaginary part of $\Sigma_\sigma$ directly encodes the lifetime of quasiparticles.
 Finite-temperature corrections vanish as $\sim 1/T$ in this limit.
The total self-energy also contains the Hartree term, $U\nu$, while all other $U^2$ diagrams cancel due to the spin structure in Eq.\eqref{ahubbard}\cite{ikeda}. The Hartree contribution can be absorbed into a shift of the Matsubara frequency, $i\omega_n\to i\omega_n-U\nu$, effectively merging with the chemical potential. Thus, Eq.~\eqref{self1} captures the leading interaction correction.

We focus on the imaginary part of the self-energy, which corresponds to the decay rate or inverse lifetime of  excitations\cite{bruus}.
In the thermodynamic limit ($N\to\infty$), the sum over $p$ becomes an integral, yielding
\begin{gather}
\text{Im}\Sigma_\sigma(k,\omega)=-\frac{U^2\nu(1-\nu)}{2\pi} \times\nonumber \\
\int_0^{2\pi} \frac{dq}{\sqrt{ \left(2J_{\bar\sigma}\sin(q/2)\right)^2-(\omega+J_\sigma \cos(k-q))^2}},
\label{ringself}
\end{gather}
where the integrand is taken to vanish whenever the expression under the square root is negative.

\paragraph{Falicov-Kimball model.}

In the Falicov-Kimball limit~\cite{falicov,zlatic,freericks,Li2019}, where one species becomes immobile 
(e.g., $J_\uparrow=0$), the self-energy simplifies significantly. It becomes momentum independent, with 
the lifetime of the immobile fermions diverging at low frequency as $\text{Im}\Sigma_\uparrow(\omega) 
\sim -\frac{U^2}{J_\downarrow}\ln|J_\downarrow/\omega|$ for $\omega\to 0$. In contrast, the mobile 
fermions exhibit a divergence at the band edge, $\text{Im}\Sigma_\downarrow(\omega) \sim -U^2/\sqrt{J_\downarrow^2-\omega^2}$, reflecting the one-dimensional density of states.

This latter behavior can be understood by noting that, at infinite temperature, the averaging over the 
immobile species effectively becomes quenched, reducing the problem to a model of binary, uncorrelated 
disorder~\cite{hodson}. Within this picture, the above self-energy expressions correspond to standard 
Born scattering results. Higher-order processes, incorporated via the self-consistent Born 
approximation, regularize these divergences, similarly to how disorder smooths the van Hove singularity 
in a clean one-dimensional chain.

\begin{figure}[ht]
\centering  
\includegraphics[width=6.5cm]{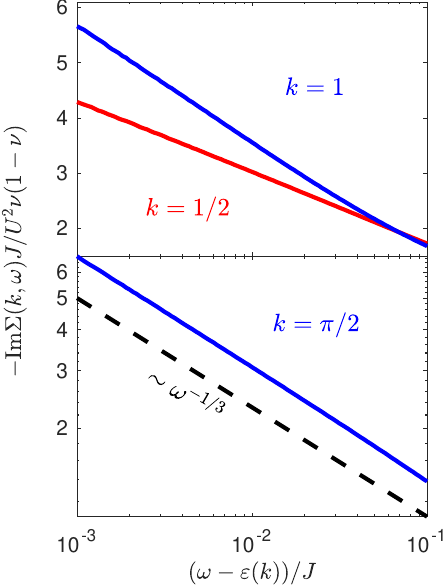}
\caption{The frequency dependence of the numerically evaluated imaginary part of the self-energy from Eq. \eqref{ringself} is shown for the symmetric Hubbard model
away from and at $k=\pi/2$ as a function of frequency (measured from the single particle energy)
on a semilog and loglog scale,
respectively. The black dashed line denotes the power law from Eq. \eqref{powerlaw}. In the upper panel, the slope changes according to $1/|\cos(k)|$, in accord with Eq. \eqref{sigmalog}.}
\label{imsigma}
\end{figure}

\paragraph{Symmetric Hubbard model.}

We now turn to the symmetric case where the two spin species have identical hoppings, $J\equiv J_\uparrow=J_\downarrow$. In this regime the self-energy develops divergences on the mass shell, $\omega=\varepsilon(k)$, whose precise nature depends strongly on both momentum and frequency. The most singular behavior occurs close to “half filling,” that is, near $\omega=0$ and $k=\pi/2$. Expanding around this point by setting $\omega\to 0$ and $k=\pi/2+\delta k$ with $\delta k\to 0$, the integrand of Eq.~\eqref{ringself} for $\omega\neq J\delta k$ takes the form  
\begin{gather}
\int_0^{2\pi}\frac{dq}{\sqrt{4J^2\sin^4(q/2)+J^2\sin(2q)\delta k-2\omega J\sin(q)}}\sim \nonumber\\
\sim \int_0^{\infty}\frac{2dq}{\sqrt{J^2q^4/4+2Jq(J\delta k-\omega)}}\sim |\omega-J\delta k|^{-1/3}.
\label{powerlaw}
\end{gather}
In the limit $J\delta k=\omega=0$, the small-$q$ contribution diverges as $1/q^2$. This divergence is cut off once the additional $2q(J\delta k-\omega)$ term is included, leading to the characteristic $-1/3$ power-law singularity. Importantly, this divergence appears only for $\omega=J\delta k=0$; when $\omega=J\delta k\neq 0$, higher-order terms under the square root prevent the singularity and convert it into a logarithmic one, as discussed below. Consequently, the self-energy close to this point behaves as  
\begin{gather}
\textmd{Im}\Sigma(k,\omega)\sim -U^2\nu(1-\nu)J^{-2/3}|\omega-J\delta k|^{-1/3}.
\label{powerlaw1}
\end{gather}
The corresponding real part also diverges, Re$\Sigma(\pi/2,\omega)\sim U^2\,\text{sgn}(\omega)|\omega|^{-1/3}$, 
consistent with the Kramers–Kr\"onig relation. This nontrivial exponent relies on the full tight-binding dispersion 
including curvature and cannot be obtained from a linearized spectrum around the Fermi points. In this respect, 
the situation resembles curvature-induced lifetime effects in Luttinger liquids at zero temperature~\cite{Samokhin}. 
Precisely at $k=\pi/2$, the imaginary part of the self-energy remains finite only for $|\omega|/J<3\sqrt{3}/2$. 
The numerically evaluated results are shown in Fig.~\ref{imsigma}. By contrast, at zero temperature the same ring diagram 
produces $\text{Im}\,\Sigma\sim U^2|\omega|$~\cite{rhodes,zlatic91,kang1994,galan}, which vanishes as $\omega\to 0$ and signals non-Fermi liquid behavior.

This anomalous divergence means that the interacting Green’s function cannot be constructed perturbatively from the noninteracting one. 
Perturbation theory breaks down when $\Sigma$ becomes comparable to $\omega$, namely when $\omega\sim U^2\omega^{-1/3}$. As a result, 
the perturbative expressions remain valid only for $\omega>U^{3/2}$, whereas at smaller frequencies the self-energy is expected to saturate 
at a scale $\sim U^{3/2}/J^{1/2}$. The divergence of the self-energy also implies that the spectral function vanishes as $|\omega|^{1/3}$ at 
$k=\pi/2$. Fourier transforming to the time domain, the momentum-resolved spectral function decays as a power law $\sim t^{-4/3}$~\cite{delft}. 
For longer times this power law crosses over to exponential decay, governed by the finite self-energy at zero frequency with decay rate $\sim U^{3/2}$. 
An analogous $-1/3$ power law divergence occurs also at $k=-\pi/2$.

Moving away from the special point $k=\pm\pi/2$, we set $\omega=-J\cos(k)+\delta\omega$ and expand the integrand in the small-$q$ limit, which now takes the form  
\begin{gather}
\int_0^\Lambda \frac{dq}{\sqrt{J^2q^2\cos^2(k)-2Jq\delta\omega\sin(k)}}\sim \frac{1}{|\cos(k)|}\ln\left(\frac{\Lambda J}{\delta\omega}\right),
\end{gather}
with $\Lambda$ a high-momentum cutoff. Exactly on the mass shell ($\delta\omega=0$), the integrand diverges as $1/q$, leading to a logarithmic divergence of the self-energy as one approaches the mass shell through either momentum or frequency,  
\begin{gather}
\textmd{Im}\Sigma(k,\omega)\sim -\frac{U^2\nu(1-\nu)}{|J\cos(k)|}\ln\left(\frac{J}{|\delta\omega-J\sin(k)\delta k|}\right),
\label{sigmalog}
\end{gather}
where $\delta k$ denotes the momentum deviation from the mass-shell condition. This result agrees with Ref.~\cite{keyser}. Unlike the imaginary part, the real part remains finite but develops a discontinuity at the mass shell of order $U^2\,\text{sgn}(\omega)$.

A consistency analysis similar to that after Eq.~\eqref{powerlaw1} applies here as well. The logarithmic divergence in Eq.~\eqref{sigmalog} is perturbatively reliable only if it remains smaller than the bare term, which requires $\delta\omega>U^2/J$. For smaller $\delta\omega$, the self-energy saturates at $\sim U^2\ln(U^2/J^2)/J$. These analytic results are illustrated in Fig.~\ref{imsigma}. As $k$ approaches $\pi/2$, the slope of the logarithmic divergence increases, signaling the crossover to the stronger $-1/3$ power-law singularity. Finally, upon departing from the symmetric limit, the behavior continuously evolves toward the Falicov–Kimball regime, where the slower species acquire a logarithmically divergent self-energy.

\paragraph{Numerics.}
\begin{figure}[t!]
\centering
\includegraphics[width=8cm]{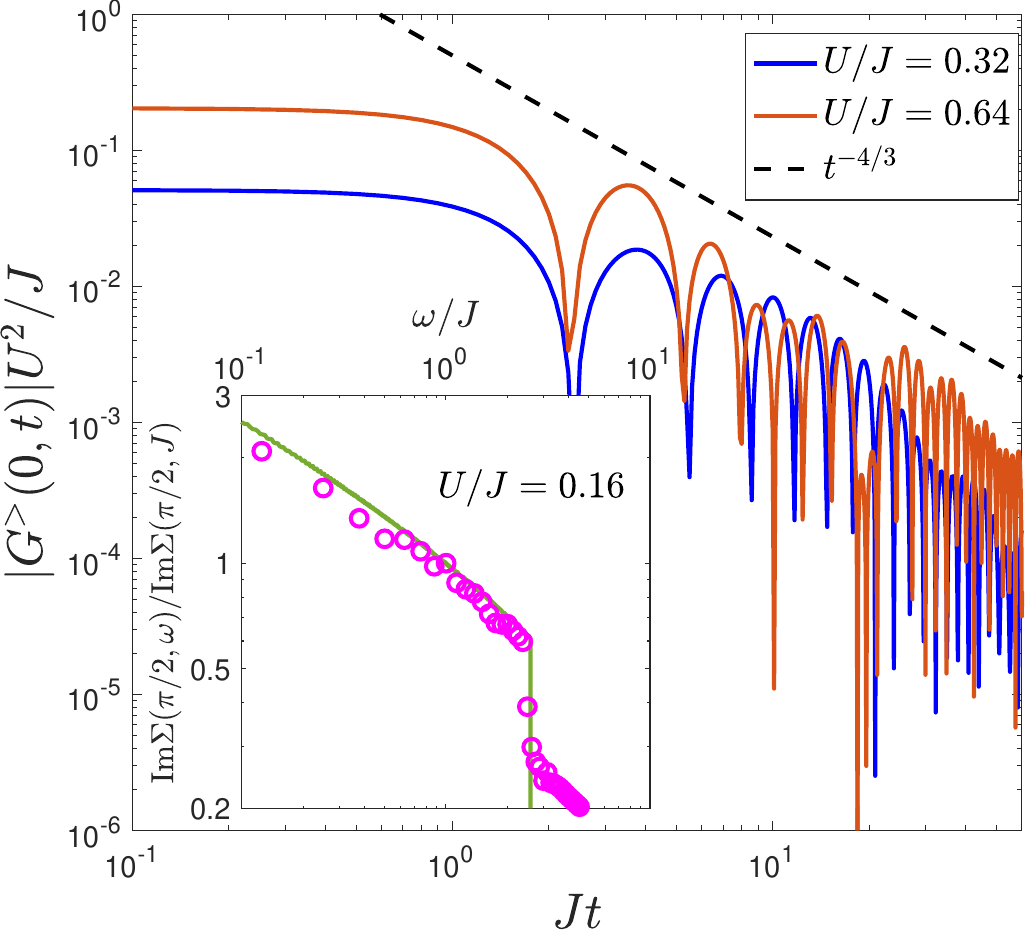}
\caption{Time evolution of $|G^>(x=0,t)|$ for the symmetric Hubbard model, displaying 
 a power law decay $\sim t^{-4/3}$ is shown for system size $N=200$. The inset shows the frequency dependent redundant self-energy (symbols) at $k=\pi/2$ for $N=100$, calculated from
the composite fermion spectral function, compared to the perturbative result from Eq. \eqref{ringself}, displaying the same $\sim \omega^{-1/3}$ frequency dependence.}
\label{fig:greendecay}
\end{figure}

To assess the validity range of the perturbative expression in Eq.~\eqref{powerlaw1}, one could in principle examine higher-order ring diagrams together with other relevant subsets of diagrams summed to infinite order, or alternatively employ numerically exact approaches that fully capture all interaction effects within numerical precision. We pursue the latter route and compute both the Green’s function and the reducible (improper) self-energy of the symmetric Hubbard model directly at infinite temperature. These quantities are obtained by propagating local operators in the Heisenberg picture while holding the density matrix fixed. At infinite temperature, the density matrix reduces to a trivial product of on-site identities, corresponding to a Matrix Product Operator (MPO) with bond dimension one~\cite{Schollwock2011}. Within this setup, the greater Green’s function~\cite{zubarev} can be written as  
\begin{gather}
i G^{>}(x,t) = \text{Tr}\left\{ \left(e^{iHt} c_{x\sigma} e^{-iHt}\right) c^\dagger_{0\sigma} \right\}.
\end{gather}
Time evolution is carried out by the time-evolving block decimation algorithm~\cite{Vidal.2004} within 
the tensor-network framework, implemented using the ITensor library~\cite{itensor}. In an analogous fashion, 
we also evaluate the propagator of the composite fermion operator $F_x = c_{x\sigma}n_{x\bar\sigma}$. For the noninteracting case, 
this correlator corresponds exactly to Fig.~\ref{ringdiag}, namely to $\Sigma(k,\omega)/U^2$, whereas for finite $U$ it yields the reducible self-energy~\cite{Fetter,bruus}. 
After Fourier transformation to momentum and frequency space, its spectral function directly provides the reducible self-energy for $k=\pi/2$ and $\omega$, which
approaches the irreducible or proper self-energy for weak interactions.

We evaluate the local Green’s function $G^>(x=0,t)$ and find that it exhibits clear power-law decay. As displayed in Fig.~\ref{fig:greendecay}, the time dependence follows $|G^>(0,t)| \sim t^{-4/3}$ over an extended temporal regime. This exponent agrees precisely with the analytic prediction obtained from the singular self-energy scaling, providing direct numerical confirmation of the 
anomalous decay mechanism at infinite temperature. Since the power-law divergent self-energy of Eq.~\eqref{powerlaw1} dominates over a finite region of $(k,\omega)$ space, the associated anomalous power law naturally manifests also in real-space observables, consistent with the numerical results above. At later times, deviations from the $t^{-4/3}$ law set in, originating both from the crossover to exponential decay governed by the finite zero-frequency self-energy and from finite-size effects (the simulations are performed with $N=200$ sites).

We also analyze the spectral function of the composite fermion operator $F_{x}$ using the same procedure. Its Fourier-transformed frequency dependence, shown in the inset of Fig.~\ref{fig:greendecay}, matches the perturbative predictions very well, clearly reproducing the $\omega^{-1/3}$ scaling of the self-energy. Additionally, the spectral weight exhibits a pronounced drop around $\omega/J\sim 3\sqrt{3}/2$, in agreement with the perturbative analysis. Together, these numerical observations corroborate the analytic results, confirming both the temporal decay of the local Green’s function and the nontrivial frequency dependence of the self-energy.

\paragraph{Conclusions.}

We have studied the self-energy of the one-dimensional Hubbard chain at infinite temperature, which 
controls the decay of quasiparticle excitations. In the asymmetric case with unequal hoppings, the 
imaginary part of the self-energy generically diverges logarithmically near the mass shell. In the 
symmetric case, this anomaly is even stronger: it develops a power-law divergence with exponent $-1/3$ 
around $k=\pm \pi/2$, signaling extremely short-lived excitations.

Tensor-network simulations confirm these predictions: the local Green’s function decays as $\sim t^{-4/3}$, 
while the composite fermion spectral function scales as $\sim \omega^{-1/3}$. These results provide direct evidence 
of anomalous relaxation in both time and frequency domains, demonstrating that quantum anomalous response and 
the breakdown of perturbation theory persist even at infinite temperature. Such effects should be 
accessible in cold-atom experiments probing dynamics and spectra in the Hubbard chain.

Future work should clarify how higher-order processes and resummations regularize these divergences, 
and investigate potential links to Kardar–Parisi–Zhang–type universality in transport and operator 
spreading. More broadly, our findings highlight that non-perturbative quantum dynamics can survive even 
in maximally hot many-body systems.

\paragraph{Note added.} During the preparation of this manuscript, we became aware of Ref. \cite{keyser}. Overlapping results are in agreement.

\begin{acknowledgments}
We thank Gergely Zar\'and for suggesting the composite fermion operator and Curt von Keyserlingk for an interesting talk which inspired us to undertake this study.
This work was supported by the National Research, Development and Innovation Office - NKFIH  Project Nos. K134437 and K142179, by a grant of the Ministry of Research, Innovation and
 Digitization, CNCS/CCCDI-UEFISCDI, under projects number
PN-IV-P1-PCE-2023-0159 and PN-IV-P1-PCE-2023-0987.
This work was also supported by the HUN-REN Hungarian Research Network through the Supported Research Groups
Programme, HUN-REN-BME-BCE Quantum Technology Research Group (TKCS-2024/34).

\end{acknowledgments}

\bibliographystyle{apsrev}
\bibliography{refgraph}

\end{document}